\documentclass[conference]{IEEEtran}
\IEEEoverridecommandlockouts
\usepackage{amsmath,amsfonts}
\usepackage{array}
\usepackage{textcomp}
\usepackage{stfloats}
\usepackage{url}
\usepackage{verbatim}
\usepackage{graphicx}
\usepackage{cite}
\usepackage{algorithm2e}
\RequirePackage{letltxmacro}
\LetLtxMacro{\LaTeXtextbf}{\textbf}
\LetLtxMacro{\textbf}{\LaTeXtextbf}
\usepackage{cite}
\usepackage{amsmath,amssymb,amsfonts}
\usepackage{graphicx}
\usepackage{textcomp}
\usepackage{caption}
\usepackage{subcaption}
\usepackage[bookmarks=false]{hyperref} 
\usepackage{algpseudocode}
\usepackage{array}
\usepackage{enumitem,lipsum}
\setlist[itemize,enumerate]{leftmargin=*}
\usepackage{etoolbox}
\usepackage{xcolor}
\usepackage{graphicx}
\usepackage{mathptmx}
\usepackage{svg}
\usepackage{listings}

\usepackage[affil-it]{authblk} 

\usepackage{tcolorbox}
\tcbuselibrary{breakable}
\usepackage{svg}

\newcommand{\etal}{\textit{et al.}}

\newtheorem{definition}{Definition}[section]

\ifCLASSINFOpdf

\else

\fi

\begin{document}

\title{VerifBFL: Leveraging zk-SNARKs for A Verifiable Blockchained Federated Learning}


\author[1,2]{Ahmed Ayoub Bellachia}
\author[1]{Mouhamed Amine Bouchiha}
\author[1]{Yacine Ghamri-Doudane}
\author[1]{Mourad Rabah}

\affil[1]{L3I, University of La Rochelle, La Rochelle, France}
\affil[2]{Ecole Nationale Supérieure d'Informatique, Algiers, Algeria}

\maketitle

\begin{abstract}

    Blockchain-based Federated Learning (FL) is an emerging decentralized machine learning paradigm that enables model training without relying on a central server. Although some BFL frameworks are considered privacy-preserving, they are still vulnerable to various attacks, including inference and model poisoning. Additionally, most of these solutions employ strong trust assumptions among all participating entities or introduce incentive mechanisms to encourage collaboration, making them susceptible to multiple security flaws. This work presents VerifBFL, a trustless, privacy-preserving, and verifiable federated learning framework that integrates blockchain technology and cryptographic protocols. By employing zero-knowledge Succinct Non-Interactive Argument of Knowledge (zk-SNARKs) and incrementally verifiable computation (IVC), VerifBFL ensures the verifiability of both local training and aggregation processes. The proofs of training and aggregation are verified on-chain, guaranteeing the integrity and auditability of each participant's contributions. To protect training data from inference attacks, VerifBFL leverages differential privacy. Finally, to demonstrate the efficiency of the proposed protocols, we built a proof of concept using emerging tools. The results show that generating proofs for local training and aggregation in VerifBFL takes less than 81s and 2s, respectively, while verifying them on-chain takes less than 0.6s. 
    

\end{abstract}


\IEEEpeerreviewmaketitle

\begin{tcolorbox}[breakable,boxrule=1pt,colframe=black,colback=white]
\scriptsize Paper accepted at IEEE/IFIP Network Operations and Management Symposium (NOMS'2025) IEEE, 2025.
\end{tcolorbox}

\section{Introduction}

\IEEEPARstart{T}{he} rapid advancement of machine learning technologies has driven remarkable progress across numerous fields, including healthcare, finance, and many other application domains. However, the success of these models often relies on access to large amounts of data, raising critical concerns about privacy and data security. To address these concerns, regulations such as the General Data Protection Regulation (GDPR) have been introduced, placing strict limits on data collection and processing practices to protect user privacy. In this context, Federated Learning (FL) \cite{mcmahan2017communication} has emerged as a promising approach to addressing these concerns by enabling decentralized model training across multiple devices without requiring raw data sharing. Despite its potential, FL comes with inherent flaws, particularly in ensuring both privacy and the integrity of contributions from participating entities. Research has shown that FL is vulnerable to various privacy attacks, including poisoning and inference attacks. \cite{qureshi2022poisoning,mothukuri2021survey,lyu2020threats}. 

\quad Blockchain, renowned for its decentralized and immutable ledger, offers complementary features that can help mitigate some of the inherent weaknesses in federated learning\cite{kim2019blockchained,nguyen2021federated}. The integration of blockchain with federated learning has led to the development of a new paradigm known as blockchain-based federated learning (BFL) \cite{8905038}. This emerging concept is driven by its potential to introduce decentralization, trustworthiness, and tamper resistance. However, due to the transparent nature of blockchain, BFL is vulnerable to free-riding attacks. Thus, considering that all participants contribute fairly and equitably is a strong assumption. For example, a ``lazy'' trainer might evade the computational burden of local training by submitting falsified updates or employing replay attacks, thus undermining the performance of the global model. Similarly, a ``lazy'' aggregator might attempt to reduce computational costs by falsifying the aggregated weights. These challenges underscore the need for robust mechanisms to enhance the integrity of BFL frameworks.

\quad To further enhance security and trustworthiness, FL is often augmented with defensive mechanisms to ensure privacy and verifiability. These mechanisms address key concerns regarding the protection of local data and the integrity of model updates, for instance, \cite{feng2024voyager} presents a Moving Target Defense (MTD) based approach designed to mitigate poisoning attacks. However, achieving verifiability and robust privacy protection in a blockchain-powered federated learning system remains a complex challenge. Privacy techniques such as differential privacy\cite{dwork2006differential}, homomorphic encryption\cite{paillier1999public}, and gradient masking \cite{bonawitz2017practical} come at a cost. Differential privacy often reduces data utility and model accuracy, the trade off is shown to be significant in critical applications \cite{nomsperformance}, homomorphic encryption introduces significant computational overhead, and gradient masking is prone to privacy leakage ~\cite{mothukuri2021survey}. Verifiability, on the other hand, is typically ensured through incentives or techniques that assess the quality of contributions. Techniques such as Multi-Krum\cite{blanchard2017machine} are employed to evaluate the quality of shared updates. Additionally, advanced cryptographic techniques, such as zk-SNARKs, have been considered in FL systems to ensure the integrity of the training process\cite{smahi2023bv},\cite{wang2023enhancing}. However, these approaches are limited to proving inference correctness and do not provide a thorough assessment of the quality of shared updates. Furthermore, the impact of such solutions on the overall performance of the underlying blockchain has not undergone a thorough analysis. Especially, in the case of a crowdsourced BFL, where training agents are supposed to engage in multiple tasks published by different task publishers. 

\quad We propose a blockchain-based crowdsourcing framework for federated learning, designed to ensure accountability and integrity throughout the FL process. To our knowledge, our contribution marks the first attempt towards using recursive zk-SNARK proofs, particularly Nova\cite{kothapalli2022nova}, to provide end-to-end verifiability of the FL workflow. At the end of each training round, every training node provides a zk-SNARK proof attesting to the accuracy of its local update. Similarly, the aggregator provides a zk-SNARK proof attesting to the integrity of the resulting global model. These proofs are then validated on the blockchain, enhancing transparency and auditability. Furthermore, we incorporate differential privacy \cite{dwork2006differential} to safeguard shared local updates from privacy attacks.

The main contributions of this paper are as follows:

\begin{itemize}
    \item We propose VerifBFL, a blockchain-based crowdsourcing framework for federated learning, where privacy preservation and end-to-end verifiability are guaranteed, to provide ground for trustless and auditable crowdsourcing.
    \item We leverage Nova's recursive argument\cite{kothapalli2022nova} to provide integrity throughout the FL process. The main idea uses recursive zk-SNARKs to construct proofs of model accuracy and global model aggregation.
    \item We present a theoretical security and privacy analysis of VerifBFL, implement a proof-of-concept and evaluate its performance. The theoretical analysis and experimental results demonstrate the efficiency of our proposed framework.
\end{itemize}

\textbf{Paper organization.} Sec.\ref{sec:prelim} describes the preliminaries and building blocks. Sec.\ref{sec:relatedWork} presents the recent advances in privacy-preserving and verifiable BFL. Sec.\ref{sec:design} presents the design of our approach. Sec.\ref{sec:proofs} details the construction of zk-SNARK proofs within VerifBFL. Sec.\ref{sec:sec-analysis} present the security analysis. Sec.\ref{sec:evaluation} shows the experimental evaluations. Sec.\ref{sec:conclusion} concludes the paper.

\section{Preliminaries}\label{sec:prelim}

\subsection{Incrementally Verifiable Computation}

IVC schemes\cite{valiant2008incrementally} allows to succinctly verify a set of iterative computations. Formally, let $\mathsf{F}$ be a function representing an arbitrary computation, $w_0, .., w_{i-1}$ be witness auxiliary inputs, and $z_0$ an initial input; define $\forall k \in [i]: z_{k+1} = \mathsf{F}(z_k,w_k)$. At each
incremental step, the IVC prover produces a proof that the step was computed correctly and it has verified a proof for the prior step. In other words, IVC allows to incrementally generate a proof $\pi_i$ claiming that there exist $w_0, .., w_{i-1}$ such that $z_i$ was correctly computed from $z_0$.

\subsection{Nova zk-SNARK}

Nova\cite{kothapalli2022nova} achieves IVC from a folding scheme for committed relaxed Rank-1 Constraint Systems (R1CS); in each iteration of IVC, the computation is folded into a running instance. Nova provides a zk-SNARK of a valid IVC proof which is composed of a tuple of algorithms \{ $\mathsf{G}$, $\mathsf{K}$, $\mathsf{P}$, $\mathsf{V}$ \} such that:
\begin{itemize}
    \item $\mathsf{G} (1^\lambda) \rightarrow (pp)$: takes as input a security parameter $\lambda$ and produces public parameters necessary for folding and SNARK proof generation.
    \item $\mathsf{K} (pp, \mathsf{F}) \rightarrow (pk, vk)$: takes as input the public parameters generated in $\mathsf{G}$ and outputs proving and verification keys for both the folding scheme and SNARK proof.
    \item $\mathsf{P} (pk, (i, z_0, zi), \Pi) \rightarrow \pi$: takes as input the proving key $pk$, the initial input $z_0$, the input of the current step $z_i$ along with the proof of correct execution of invocations $0,..,i-1$ of $F$ and produces a proof $\pi$ that $\Pi$ is valid and $z_{i+1} = \mathsf{F}(z_i)$. 

    \item $\mathsf{V} (vk, (i, z_0, z_i), \pi) \rightarrow \{ 0, 1\}$: takes as input the verification, the initial input $z_0$, the output $z_i$ and the IVC proof $\pi$, outputs $1$ if $\pi$ is valid and $0$ otherwise.
    
\end{itemize}

\subsection{Differential Privacy}

Differential privacy (DP) guarantees the privacy and utility of data with rigorous theoretical foundation\cite{dwork2006differential}. The intuition behind DP is to add noise that follows a carefully chosen probability distribution to the original data without compromising its utility. This guarantees the privacy of data regardless of any and all sources of background information. Differential privacy was first introduced in\cite{dwork2006differential} and given the following formal definition: 

\begin{definition}[$\epsilon$-Differential Privacy]
A randomized mechanism $M$ with domain $\mathbb{N}^{|\chi|}$ and range $R$ satisfies $\epsilon\text{-differential privacy}$ if for every adjacent datasets $D, D' \in \mathbb{N}^{|\chi|}$ and any subset $S \subseteq R$ we have:
\begin{center}
    $Pr[M(D) \in S] \leq e^\epsilon Pr[M(D') \in S]$, where $\epsilon \ge 0$ is a privacy parameter.
\end{center}
\end{definition}

\section{Related work} \label{sec:relatedWork}

The combination of blockchain with federated learning gained a significant interest in different domains, where the main focus is to provide ground for privacy-preserving and verifiable federated learning, the proposed BFL frameworks, often rely on mechanisms to meet the security requirements of real-world applications. In \cite{zhao2020privacy}, a BFL framework for IoT devices uses differential privacy to protect local updates but relies solely on multi-Krum to filter anomalies, which, besides its computational overhead, overlooks replay attacks and double-spending. Moreover, its reputation-based incentive mechanism may lead participants to prioritize quantity over quality.

\quad In \cite{fan2020hybrid}, a hybrid blockchain-based system for resource trading in federated learning (FL) within the context of edge computing is proposed. This system combines public and private blockchains to achieve a balance between transparency and efficiency. To mitigate malicious updates from clients, the authors incorporate detection techniques like Reject on Negative Influence (RONI) \cite{barreno2010security} and Foolsgold to filter out poisoned model updates. However, the framework lacks privacy protection for shared updates, and AI-driven detection adds computational overhead, especially in cross-device FL scenarios. 

\quad In\cite{wang2022blockchain}, the authors present a BFL framework for the Internet of Vehicles (IoV) where the shared local updates are encrypted before being shared using the Paillier scheme\cite{paillier1999public}. The scheme is leveraged to perform global model aggregation on encrypted gradients; safeguarding from curious aggregators. However, this comes at the expense of huge computational overhead. Additionally, with the use of multi-Krum as the only means to assess the participants' contribution, the system is vulnerable to deviating entities.

\quad In \cite{zhou2024vdfchain}, the authors leverage the gradient masking technique introduced in \cite{bonawitz2017practical} to protect against data leakage during model updates. Furthermore, they employ polynomial commitment schemes to ensure the verifiability of the aggregation process and to mitigate Byzantine behaviors. However, the system assumes that all training nodes act honestly, which is an unrealistic assumption under real-world conditions. Additionally, The use of the gradient masking in \cite{bonawitz2017practical}, coupled with the use of polynomial commitments, introduces significant computational overhead, particularly in cross-device federated learning setups where resource constraints are more pronounced.

\quad The use of Zero-Knowledge Proofs (ZKP) to provide verifiable computation has been recently considered in the context of machine learning. Notably, zkFL \cite{wang2024zkfl} uses zero-knowledge proofs to guarantee the integrity of the aggregation process. To attest to correct aggregation results, the aggregator provides a proof per round that demonstrates to clients that the aggregator faithfully executes the expected behavior. However, this verifiability is achieved in a centralized setup and does not cover the training process, leaving it vulnerable to poisoning attacks and colluding parties.  Smahi \etal, \cite{smahi2023bv}, propose a BFL framework for Vehicle-to-Everything (V2X) environments, where Local Differential Privacy (LDP) is employed for privacy preservation. Unlike classical Differential Privacy, which adds noise to the gradients, LDP adds noise directly at the data or input level. This differentially private data is then used to train a Support Vector Machine (SVM) model. The authors rely on a variant of zk-SNARKs called CP-SNARK, which is employed by the training clients to provide proofs of correctness for their shared updates. This work marks the first instance of integrating this cryptographic primitive within a decentralized setup. However, the proof generation, as presented in the article, only proves one inference operation and does not prove the full training process, hence, it does not provide sufficient information about the quality of the prover's contribution. 


\quad In the work presented in\cite{wang2023enhancing}, only CRT is relied upon as means to mask local updates since it allows for a more direct aggregation process. Additionally, the authors adopt zk-SNARKs with an accelerated version of the Groth16\cite{groth2016size}, one of the earliest zk-SNARK constructions. At the end of each training round, training clients provide a proof of the integrity of the generated local models. The authors evaluate the use of this primitive to prove the training of a convolutional neural network (CNN). However, the proof does not attest to the whole training process, rather; only inference is proved. 


\quad Overall, existing solutions ensure privacy, but often at the cost of data utility or significant computational overhead. Moreover, many systems fail to accurately assess the contributions of their participants, leaving them vulnerable to malicious behavior. The evaluation of recent work underscores the need for more efficient and resilient mechanisms, as well as a notable research gap towards achieving end-to-end verifiability in BFL which we aim to address with our approach.

\section{System Design}\label{sec:design}

\begin{figure}[t]
    \centering
    \includegraphics[width=1 \linewidth]{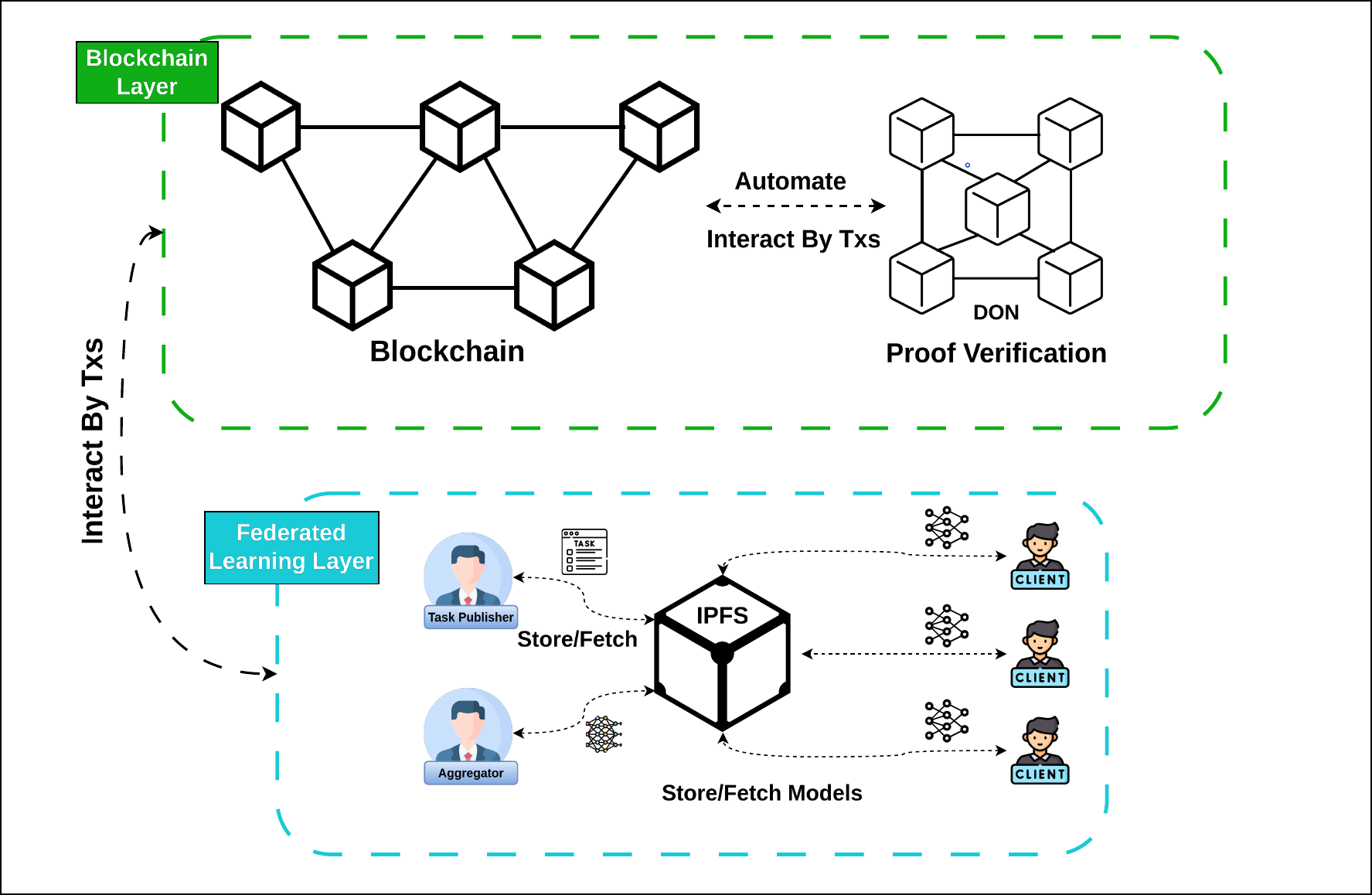}
    \caption{Design Overview: VerifBFL is composed of two layers, with the blockchain layer overseeing the work of the FL layer. TPs, TAs, and AGs store FL tasks metadata on IPFS, then submit Txs to the blockchain in a crowdsourced manner. Proofs for training and aggregation are generated off-chain by TAs and AGs and verified on-chain.}
    \label{fig:design}
\end{figure}

\subsection{System Model}
We present a blockchain-based crowdsourcing framework as depicted in Fig. \ref{fig:design} to manage federated learning tasks. VerifBFL comprises the following entities:

\begin{itemize}
    \item \textbf{Task Publishers.} Identified by $TP$, are responsible for publishing new learning tasks. A learning task $t_i$ states a description of the task along with an initial model and learning requirements. Task publishers are required to pay for the training so they deposit a reward.
    \item \textbf{Clients.} Also referred to as Trainers, identified by $TR$, participate in the training of federated learning models. They can subscribe to existing tasks and receive rewards for their contributions.
    \item \textbf{Aggregators.} Identified by $AG$, can subscribe to existing tasks and are responsible of performing model aggregation. Similarly, they receive rewards for their honest contributions.
\end{itemize}

We adopt a loosely-coupled blockchain-based federated learning setup \cite{wang2021blockchain}, where the aforementioned entities and blockchain nodes operate in separate networks. By adopting a permissioned blockchain network, we monitor and evaluate the contribution of each entity.

\subsection{Threat Model}

In this section, we present a threat model that groups the common threats that VerifBFL is faced with.

 We adopt a zero-trust stance towards all federated learning (FL) participants, meaning no actor is trusted by default to adhere to the protocol. Furthermore, we account for the possibility that participants may attempt to infer the private states of others, necessitating robust privacy-preserving and verification mechanisms. Additionally, we assume that at least $2/3$ of the blockchain's nodes behave honestly. Similarly, we expect more than $2/3$ of Decentralized Oracle Network (DON) nodes to follow the protocol, as they are responsible for verifying training and aggregation proofs. 

We summarize the potential threats to our blockchain-based federated learning as follows:

\begin{itemize}
    \item \textbf{Poisoning attacks.} This type of attacks occurs when a FL participant tries to intentionally manipulate the training data or local model updates to degrade the performance of the global model, or aims to induce the FL model to output the target label specified by them. With the latter being more diﬀicult to realize\cite{lyu2020threats}.
    \item  \textbf{Inference attacks.} The gradients exchange during the FL protocol can leads to privacy leakages\cite{huang2011adversarial}, with many types of inference attacks, such as membership inference, and class membership inference, aiming to recover sensitive attributes or properties about the training data.
    \item  \textbf{Reconstruction attacks.} Reconstruction attacks in federated learning (FL) pose a significant threat to participant privacy. In this type of attack, a malicious actor attempts to reconstruct sensitive data from the model updates shared by FL participants.
    \item  \textbf{Free-riding.} T the FL actors may attempt to reduce their computational cost to boost productivity, especially when rewards are tied to training contributions. One way they can do this is by resubmitting the same model update across multiple training rounds or making minor tweaks to previous updates that violate the FL protocol. Such replay attacks degrade the performance of the aggregated model, undermining the overall quality and reliability of the learning process.
    \item \textbf{Collusion attacks.} Collusion attacks occur when a group of nodes cooperate in a malicious manner to subvert the system’s integrity, privacy, or accuracy. In BFL systems, where participants are supposed act independently and without prior knowledge of each other's data or intentions\cite{xiao2022sca}, this type of attacks can enable adversaries to bypass some verification strategies or manipulate the behavior of the resulting global model.
\end{itemize}

Under these threats, our proposed approach aims to provide the following properties:
\begin{enumerate}
    \item \textbf{Privacy-Preservation.} In federated learning, privacy preservation is crucial to prevent the leakage of sensitive training data. we employ differential privacy as a mechanism to inhibit privacy attacks and protect the private state of training agents.
    \item \textbf{Verifiability.} According to \cite{kairouz2021advances}, verifiability in FL is defined as the ability of a training node or aggregator to prove to others participating in the FL protocol that it has executed the desired behavior faithfully, without revealing the potentially private data upon which they were acting. We aim to ensure verifiability within our system by leveraging recursive zk-SNARKs.
    \item \textbf{Accountability.} Accountability is critical in federated learning, especially in crowdsourcing scenarios with untrusted participants. We aim to achieve accountability by implementing verifiability mechanisms alongside an insensitive protocol that rewards honest behavior and penalizes deviations. 

\end{enumerate}

\subsection{VerifBFL Arichitecture}

VerifBFL consists of a loosely-coupled blockchain-based federated learning (BFL) system, in which the blockchain nodes and the FL actors function as independent entities. The blockchain layer, comprising the distributed ledger and the decentralized oracle network (DON), serves as a trusted orchestrator, overseeing the federated learning protocol. The latter occurs within the FL layer, which is comprised of three main actors, trainers/clients, aggregators, and task publishers. The Inter-Planetary File System (IPFS) is used as an off-chain storage system for the gradients exchanged between these actors. This helps mitigate the storage limitations inherent in the blockchain. Details on each of these components are provided in the following.

\subsubsection{IPFS}
Inter-Planetary File System (IPFS)\cite{benet2014ipfs} is a peer-to-peer distributed file system that enables distributed computing devices to connect with the same file system. We implement the off-chain storage using IPFS, and store hashes of data locations on the blockchain instead of actual files. This enables for easier data management within our framework and overcomes the storage limitations present in blockchain, which, for security purposes, does not support floating point types, which are heavily expressed in machine learning models. Task publishers post their task description along with the required starting parameters in the IPFS, and only a hash referring its location is stored on-chain. Similarly, FL actors use IPFS as a storage solution for local and global models that are generated throughout the FL protocol.

\subsubsection{Blockchain}
 It is responsible for managing learning tasks and ensuring an honest federated learning protocol via smart contract.
For VerifBFL, we leverage a permissioned blockchain model, where a committee of nodes takes charge of validating transactions and appending new blocks to the ledger. The blockchain network in VerifBFL employs the Practical Byzantine Fault Tolerance (PBFT) \cite{castro1999practical} consensus protocol to ensure robustness and efficiency. Furthermore, PBFT is known for its faster transaction finality, which is of particular importance in our crowdsourced FL business logic. This is because it is essential to guarantee that training agents commence a new round utilizing an identical global model. Consequently, when a global model for the current round is added to the ledger, it is regarded as definitive thanks to the byzantine agreement.

\begin{figure}[t]
    \centering
    \includegraphics[width=1 \linewidth]{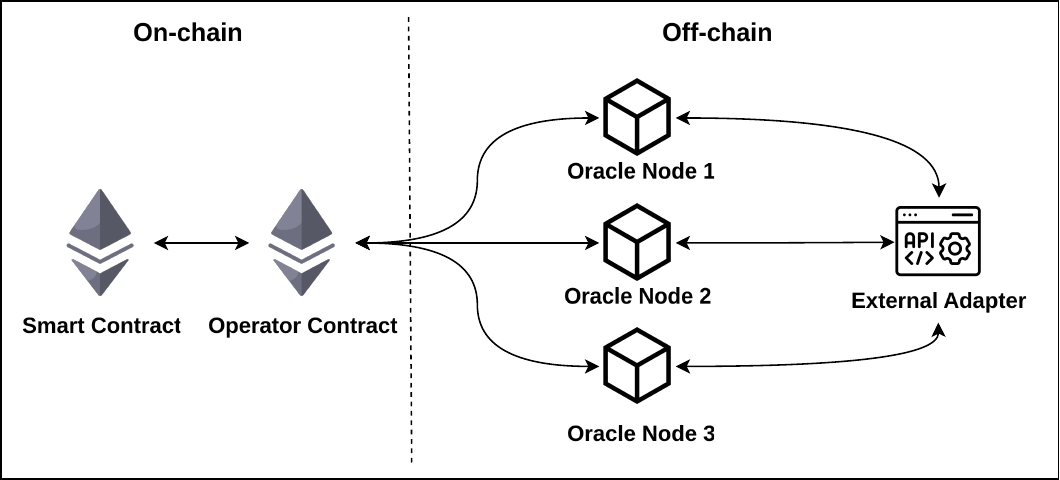}
    \caption{Chainlink Request Model: the consumer contract sends requests to Chainlink nodes through the operator contract, custom computations are performed by leveraging external adapters.}
    \label{fig:chainlink-ea}
\end{figure}

\subsubsection{Decentralized Oracle Network}
Decentralized oracle networks are designed to enhance and alleviate the performance limitations of blockchain systems by providing networking, storage and computation services while ensuring confidentiality, availability, and integrity.
DONs are formed by groups of Oracle nodes cooperating to accomplish jobs in a blockchain-agnostic manner, they do so by the intermediary of adapters\cite{breidenbach2021chainlink}. 
In our framework, we utilize Chainlink's DON to offload proof verification from the blockchain, maintaining the same level of trustworthiness and integrity while optimizing gas usage. This approach effectively overcomes the computational and memory constraints imposed within the Ethereum Virtual Machines (EVMs). Fig.\ref{fig:chainlink-ea} illustrates Chainlink's request model: through an on-chain operator contract, a consumer contract can create and send requests to Chainlink nodes for custom computations. These computations are executed by leveraging external adapters which expose an API endpoint to perform the specified tasks.

\subsection{Workflow}
In this section, we present the workflow of our crowdsourced BFL. We explain the process of training a task, from its submission to the blockchain to its accomplishment. The steps of the workflow are illustrated in Fig.\ref{fig:workflow}.

\begin{enumerate}
    \item \textbf{Task submission.} (steps 1-3 in Fig \ref{fig:workflow}) Task publishers publish learning tasks to the blockchain, a task is identified by a unique ID, referring to its location in IPFS, it comprises a description of the training task, along with an initial model and learning parameters. Additionally, task publishers specify a target accuracy that they wish to achieve along with the number of trainers and the number of training rounds. A reward that will be distributed to the actors is deposited into an escrow smart contract (ESC) by the task publisher upon task submission. The details are shown below:
    
\begin{tcolorbox}[breakable,boxrule=1pt,colframe=gray,colback=white]
\begin{small}
        
    \section*{Task Submission}

    \begin{itemize}
        \item \textbf{Input:}  Task info $t_i$ and reward $r$
        \item \textbf{Output:} Submits $t_i$ to the blockchain 
    \end{itemize}

    \begin{enumerate}
    
    \item Check that $t_i$ does not already exist
    \BlankLine
    \item Check that $r \geq R_{min}$
    \BlankLine
    \item Transfer $r$ from the $TP$ account to the $ESC$
    \BlankLine
    \item Add $t_i$ to list of Tasks
    \BlankLine
    \item Emit event \texttt{TaskSubmitted($t_i.taskId$)}
    \end{enumerate}
    \end{small}
\end{tcolorbox}

    \begin{figure}[t]
    \centering
\includegraphics[width=0.5\textwidth]{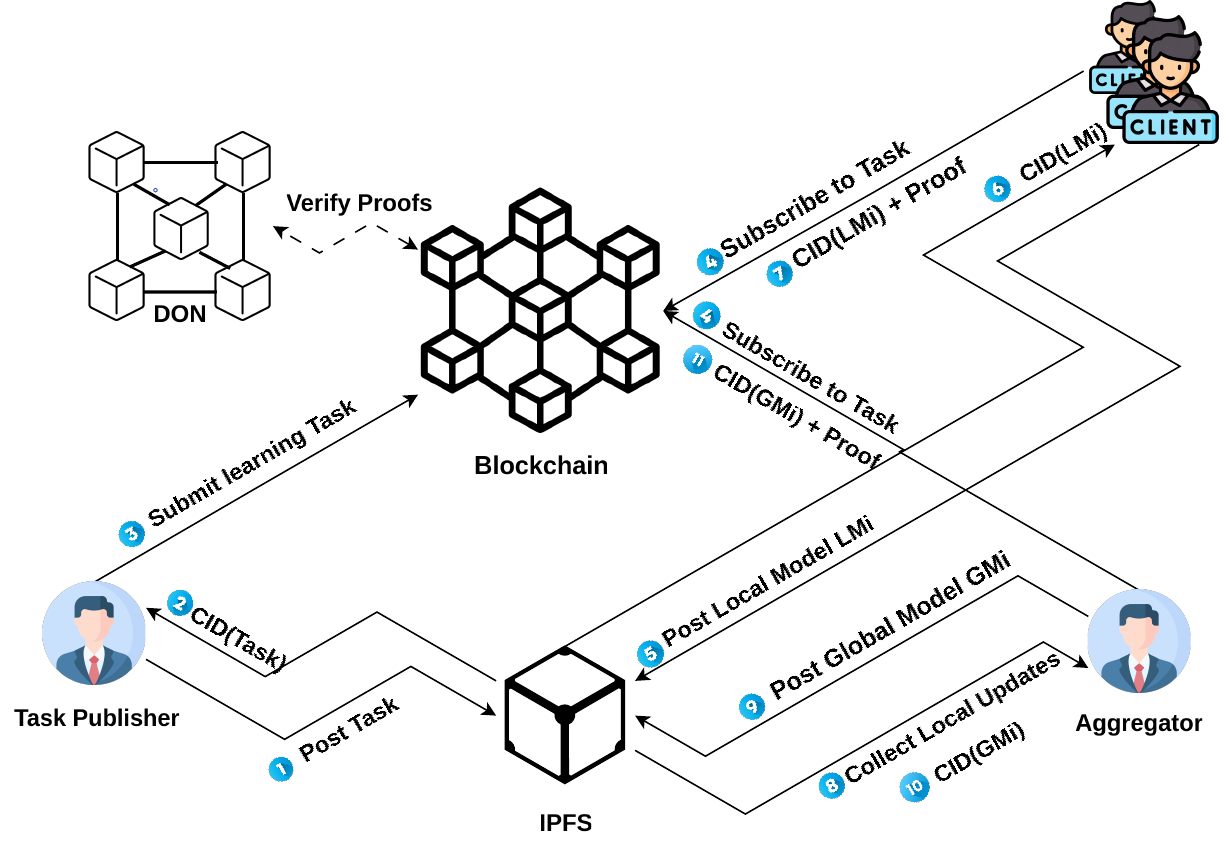}
    \caption{VerifBFL Workflow}
    \label{fig:workflow}
\end{figure}

\item \textbf{Task subscription.} (step 4 in Fig \ref{fig:workflow}) Trainers and aggregators can subscribe to available tasks by staking a specified amount of tokens as a guarantee of their honest participation. Any deviation from the federated learning (FL) protocol results in the loss of the staked amount and blacklisting of the offending entity. This staking process adds a layer of robustness, enforcing good behavior through economic incentives and penalties, while also reducing the risk of denial-of-service (DoS) attacks. Details about the subscription are provided below:
\begin{tcolorbox}[breakable,boxrule=1pt,colframe=gray,colback=white]
\begin{small}

    \section*{Task Subscription}

    \begin{itemize}
        \item \textbf{Input:}  Stake $k$ and TaskId $t_i.taskId$
        \item \textbf{Output:} Adds trainer $TR$ (\text{resp. aggregator }$AG$) to list of trainers $t_i.trainers$ (\text{resp. aggregators} $t_i.aggregators$) 
    \end{itemize}

    \begin{enumerate}
    
    \item Check that $TR \notin t_i.trainers$ (\text{resp. }$AG \notin t_i.aggregators$) 
    \BlankLine
    \item Check that $k \geq K_{min}$
    \BlankLine
    \item Lock the staked amount $k$ in $ESC$
    \BlankLine
    \item Emit event \texttt{Subscribed} ($t_i.taskId$, $TR$.adr (resp. $AG$.adr))
    \end{enumerate}
    \end{small}
\end{tcolorbox}
    
    \item \textbf{Local Training.} (steps 5-7 in Fig \ref{fig:workflow}) At the beginning of every global training round, clients/trainers are invited to download the global model from IPFS, and perform local training using their own dataset, the resulting local models are injected with $\epsilon\text{-differentially private}$ noise to protect them from privacy attacks. Subsequently, clients generate a zk-SNARK proof $\pi$ of the integrity of the generated model. The local model is stored in IPFS, with a hash referring to it submitted to the blockchain along with the proof $\pi$ for verification. 

    \item \textbf{Aggregation.} (steps 8-11 in Fig \ref{fig:workflow}) Upon local model submission, training proofs are verified in the blockchain via the decentralized oracle network (DON). Clients who submit invalid proofs are blacklisted and forfeit their staked tokens, whereas valid local models are transmitted to the aggregator to construct the global model. The aggregator downloads the local models from IPFS, performs aggregation, and generates a proof $\pi$ of the integrity of the resulting global model. The proof along with a hash referring to the global model is submitted to the blockchain for verification. If the aggregator fails to provide valid proof, they lose their share and get blacklisted; and another aggregator takes over.

    \item \textbf{Reward Distribution.} The local training and aggregation steps are repeated until the desired target accuracy is achieved or the total number of training rounds is reached. At the end of the FL task, each honest participant retrieves their stake and is rewarded for their contribution. If the resulting model does not fulfill the requirements specified by the task publisher, the latter receives a fixed amount of tokens as compensation.
\end{enumerate}

\section{zk-SNARK Proofs Construction}
\label{sec:proofs}

In this section, we outline the proof construction process within our framework. After completing local training, each client generates a proof attesting to the accuracy and integrity of their local update. Similarly, at the end of each global training round, the aggregator generates a proof verifying the correct construction of the global model. Below, we explain the rationale for using Nova \cite{kothapalli2022nova} to build both the local accuracy proofs and the global aggregation proofs.

\subsection{Proof of Accuracy}
The trainers, upon local training, generate proofs to attest the accuracy of their trained model. The statement that a prover claims is that their local model $M$ has a given accuracy $Acc$. To construct such proof, we express the accuracy computation as an incrementally verifiable computation (IVC) as introduced in the previous section. The intuition is presented in the following:
\begin{itemize}
    \item Let $F$ represent the execution of one inference operation of our model. Notably, an invocation of $F$ refers to the computation of a model prediction on an input.
    \item In our construction $z_i$ represents a counter for the number of correct predictions accomplished in the set of  invocations $\{0,1,...,i-1\}$ of $F$.
    \item Finally, at invocation $n$, $F^{(n)}$ represents the execution of n invocations of $F$, i.e, the inference of $n$ inputs to our Model. Consequently, $z_n$ will hold the number of total correct prediction of our Model. The accuracy of the model can be obtained as follows: 
     \begin{equation}
         Acc = \frac{\text{number of correct predictions}}{\text{number of total predictions}} = \frac{z_n}{n}
     \end{equation}
\end{itemize}

\begin{tcolorbox}[breakable,boxrule=1pt,colframe=gray,colback=white]
\begin{small}

\section*{Proof of Accuracy}

\begin{itemize}
    \item \textbf{Input:}  Test dataset $\{ w_0, .., w_{n-1}\}$ and public parameters $pp$.
    \item \textbf{Output:} Proof of accuracy $\pi$ 
\end{itemize}

\begin{enumerate}

\item Let $Circuits$ = $\{ c_i : c_i = InferenceCircuit(w_i) \And i \in [0,n-1]\}$
\BlankLine
\item let $RecursiveSnark$ = $\text{new RecursiveSnark}( pp, c_0, z_0 )$ 
\BlankLine
\item For $i$ in $1..n$: $RecursiveSnark.prove-step(pp, c_i)$
\BlankLine
\item Let $(pk, vk) = \mathsf{K(pp, c_0)}$
\BlankLine
\item $\pi$ = $\text{new CompressedSnark} (pk, RecursiveSnark)$
\end{enumerate}
\end{small}
\end{tcolorbox}

\subsection{Proof of Aggregation}
The aggregators, at the end of each global round, aggregate the collected local updates to obtain a global model. They provide a proof of the integrity of the obtained global model. Concretely, we implement a proof for the FedAvg algorithm \cite{mcmahan2017communication}, the prover claims the following statement: $w_{glob} \leftarrow \sum_{k \in S} \frac{n_k}{n} w_k$, where $w_{glob}$ is the global model, $w_k$ is the local model of client $k$ and $\frac{n_k}{n}$ is a weight factor equal to the ratio of data volume of client $k$ to the total data volume.

The proof is constructed as follows. For the sake of simplicity, we remove the weight factor $\frac{n_k}{n}$:
\begin{itemize}
    \item Suppose we aggregate $n$ local models $\{w_0, w_1, ..., w_n\}$, let $F$ represent the incremental addition of local models.
    \item At invocation $i$, $F_i$ represents the addition of $\{w_0, w_1, ..., w_i\}$. 
    \item At invocation $n$, we obtain the global model by adding all local models $\{w_0, w_1, ..., w_n \}$
    \item Due to implementation specifications, in this setup, the running input to $F$ : $z_i$ will be set to $1$ if invocation $i$ is performed correctly. $z_i$ cannot hold the addition result of local models because of its high dimension.
    
\end{itemize}

\begin{tcolorbox}[breakable,boxrule=1pt,colframe=gray,colback=white]
\begin{small}

\section*{Proof of Aggregation}

\begin{itemize}
    \item \textbf{Input:}  Local updates $\{ w_0, .., w_{n-1}\}$ and public parameters $pp$.
    \item \textbf{Output:} Proof of aggregation $\pi$ 
\end{itemize}

\begin{enumerate}

\item Let $Circuits$ = $\{ c_i : c_i = AggregationCircuit(w_i) \And i \in [0,n-1]\}$
\BlankLine
\item let $RecursiveSnark$ = $\text{new RecursiveSnark}( pp, c_0, z_0 )$ 
\BlankLine
\item For $i$ in $1..n$: $RecursiveSnark.prove-step(pp, c_i)$
\BlankLine
\item Let $(pk, vk) = \mathsf{K(pp, c_0)}$
\BlankLine
\item $\pi$ = $\text{new CompressedSnark} (pk, RecursiveSnark)$
\end{enumerate}
\end{small}
\end{tcolorbox}


The above proof constructions provide the following properties:
\begin{itemize}
    \item \textbf{Completeness:} an honest prover with a valid proof $\pi$ should convince the verifier. Formally, for any probabilistic poly(n)-time (PPT) adversary $\mathsf{A}$, we have: 
    \[
    \Pr \left[ 
    \mathsf{V}(vk, u, \pi) = 1 
    \middle| 
    \begin{array}{l}
    pp \leftarrow \mathsf{G}(1^\lambda), \\
    (F, (u, w)) \leftarrow \mathsf{A}(pp), \\
    (pp, F, u, w) \in \mathsf{R}, \\
    (pk, vk) \leftarrow \mathsf{K}(pp, F), \\
    \pi \leftarrow \mathsf{P} (pk, (i, z_0, zi), \Pi)
    \end{array} 
    \right] = 1.
    \]
    \item \textbf{Knowledge Soundness:} The prover cannot forge a valid proof $\pi$ without knowledge of a valid witness $w$. Formally, for all PPT adversaries $\mathsf{A}$, there exists a PPT extractor $\varepsilon$ such that for all randomness $\rho$ we have:
        \[
        \Pr \left[ 
        \begin{array}{l}
        \mathsf{V}(vk, u, \pi) = 1, \\
        (pp, F, u, w) \notin \mathsf{R}
        \end{array} 
        \middle| 
        \begin{array}{l}
        pp \leftarrow \mathsf{G}(1^\lambda), \\
        (F, u, \pi) \leftarrow \mathsf{A}(pp; \rho), \\
        (pk, vk) \leftarrow \mathsf{K}(pp, F), \\
        w \leftarrow \varepsilon(pp, \rho)
        \end{array} 
        \right] = \text{negl}(\lambda).
        \]
    \item \textbf{Succinctness:} the verifier runs in constant time $O(1)$ and the proof size is constant in the size of the step circuit that represents the computation $F$ regardless of the number of invocations of $F$. 
    \item \textbf{zero-knowledge:} The proof does not leak any information besides the truth of the statement proven. There exists a simulator $S$ such that for all PPT adversaries $\mathsf{A}$, we have:

   {\scriptsize
    \[
    \left\{
    (pp, F, u, \pi) 
    \middle|
    \begin{array}{l}
    pp \leftarrow \mathsf{G}(1^\lambda), \\
    (F, (u, w)) \leftarrow \mathsf{A}(pp), \\
    (pp, F, u, w) \in \mathsf{R}, \\
    (pk, vk) \leftarrow \mathsf{K}(pp, F), \\
    \pi \leftarrow \mathsf{P}(pk, u, w)
    \end{array} 
    \right\}
    \cong
    \left\{
    (pp, F, u, \pi) 
    \middle|
    \begin{array}{l}
    (pp, \tau) \leftarrow \mathsf{S}(1^\lambda), \\
    (F, (u, w)) \leftarrow \mathsf{A}(pp), \\
    (pp, F, u, w) \in \mathsf{R}, \\
    (pk, vk) \leftarrow \mathsf{K}(pp, F), \\
    \pi \leftarrow \mathsf{S}(pp, u, \tau)
    \end{array} 
    \right\}
    \] 
    }

\end{itemize}

These properties are obtained from the Nova proof construction \cite{kothapalli2022nova} under the discrete logarithm (DLOG) hardness assumption and provided that we heuristically instantiate the random oracle with a concrete hash function in the standard model.

\section{Security Analysis}\label{sec:sec-analysis}
In this section, we give a thorough security analysis of our proposed system by putting it under the scope of our proposed threat model. We detail the aspects that make our framework resistant to BFL threats.

\begin{enumerate}
    \item \textbf{Privacy Attacks.} Our proposed systems inhibits privacy attacks by leveraging differential privacy, as the noise added to each model update makes it difficult for an adversary to distinguish between the updates of different trainers. 
    Let $\mathsf{A}$ be an adversary attempting to perform an inference attack on our system. We have: $\mathsf{Pr}[\textit{correctGuess}]= \frac{e^\epsilon}{e^\epsilon +1}$.Thus, by setting a sufficiently low value of $\epsilon$, we can make it statistically infeasible for $\mathsf{A}$ to perform a successful inference.
    
    \item \textbf{Free Riding Attacks.} By submitting training proofs at the end of each local training round, trainers commit to the training data used. Under the DLOG hardness assumption, the knowledge soundness property of Nova ensures that the proofs cannot be forged unless the trainers have conducted the training with integrity. Moreover, these proofs allow for a direct assessment of the accuracy of the provided updates while maintaining the confidentiality of the training data, thanks to the statistical zero-knowledge property. 
    We effectively prevent poisoning attacks and eliminate the risk of free-riding by deterring lazy participants. Similarly, aggregators provide proof of the correctness of the global model generated at the end of each round, ensuring that it adheres to the aggregation strategy defined within the FL protocol. 
    \item \textbf{Collusion Attacks.} Our underlying blockchain platform makes use of the PBFT consensus to eliminate the impact of colluding nodes. Provided that the majority ($2/3$) of decentralized oracle nodes perform verification honestly, the combination of zk-SNARK proofs with the blockchain limits the extent to which colluding entities may impact the behavior of our proposed framework.
\end{enumerate}

\section{Performance Analysis} \label{sec:evaluation} 

This section present the performance evaluation of our proposed approach, we conduct benchmark tests to highlight the efficiency of our cryptographic proofs and the availability of the VerifBFL framework.

\subsection{Environment} 
We conducted the experimental tests on a Dell XPS 15 pc powered by an Intel Core i7-10750H Hexa Core CPU with a base frequency of 2.60 GHz and 16 GB of RAM.

\subsection{zk-SNARK Proofs}
\subsubsection{Experimental Setup} 
We generate proofs of training and aggregation for a convolutional neural network (CNN) using the MNIST dataset\cite{LeCun2005TheMD}. To implement these proofs, we utilize the nova-snark Rust crate\footnote{\href{https://crates.io/crates/nova-snark}{https://crates.io/crates/nova-snark}}. The proof of accuracy is verified by evaluating the model against a set of 100 images. Similarly, we produce an aggregation proof for five local models based on the FedAvg algorithm. In total, we generated 30 training and aggregation proofs and calculated the average setup, proving, and verification times, which are detailed in the following table:

\begin{table}[h]
\centering
\begin{tabular}{| c| c| c| c| c|}
\hline
\textbf{Proof}  &\textbf{Setup} & \textbf{Prove} & \textbf{Verify} & \textbf{Proof Size}\\ \hline
Training proof & 785.53s & 81.95s & 0.642s & 33.3Kb\\ \hline
Aggregation proof & 105.64s & 2.189s & 0.536s & 35.4Kb\\ \hline

\end{tabular}
\caption{Proofs Benchmark}
\label{tab:proof-bench}
\end{table}

\subsubsection{Performance Evaluation}
The obtained results shown in Tab.\ref{tab:proof-bench} highlight the efficiency of proofs generation (81s and 2s) and verification(0.6s). With this very short verification time, the on-chain computational overhead introduced to our VerifBFL framework is minimal. Additionally, the setup time, which is influenced by the size of the circuits, can be largely reduced in a multi-threaded environment and is only required at the beginning of a learning task, making its impact on the overall workflow of the crowdsourcing framework negligible.

\subsection{Crowdsourcing System}
\subsubsection{Experimental Setup} 
To assess the on-chain performance of our FL crowdsourcing framework, we use HyperLedger Caliper\footnote{\href{https://github.com/hyperledger/caliper-benchmarks}{https://github.com/hyperledger/caliper-benchmarks}}, as a performance benchmarking tool. We provide a fine-grained evaluation by testing each function of our task management smart contract individually. 

We utilize Geth\footnote{\href{https://geth.ethereum.org/}{https://geth.ethereum.org/}} to establish our EVM-based blockchain network, which is comprised of four nodes operating under the PBFT consensus mechanism. Tab. \ref{tab:benchconfig} provides a detailed summary of the network’s characteristics, including node specifications and consensus parameters.

\begin{table}[h]
\centering
\begin{tabular}{| c| c| }
\hline
Consensus & PBFT \\ \hline
Number of nodes  & 4 \\ \hline
Blockperiodseconds  & 1 \\ \hline
Timestamp & 0x58ee40ba \\ \hline
GasLimit  & 0x1fffffffffffff \\ \hline
Epochlength   & 30000 \\ \hline
\end{tabular}
\caption{Blockchain Configuration}
\label{tab:benchconfig}
\end{table}

\subsubsection{Performance Evaluation}

To assess the performance of our system, we consider two metrics for our evaluation:
\begin{itemize}
    \item \textbf{Throughput.} is expressed by the number of successful transactions per time unit.
    \item \textbf{Latency.}  refers to the elapsed time between a transaction submission and its completion.
\end{itemize}

\begin{figure*}[t]
    \centering
    \hspace{-0.2in}
    \subfloat[Task Creation]{
        \includegraphics[scale=0.23]{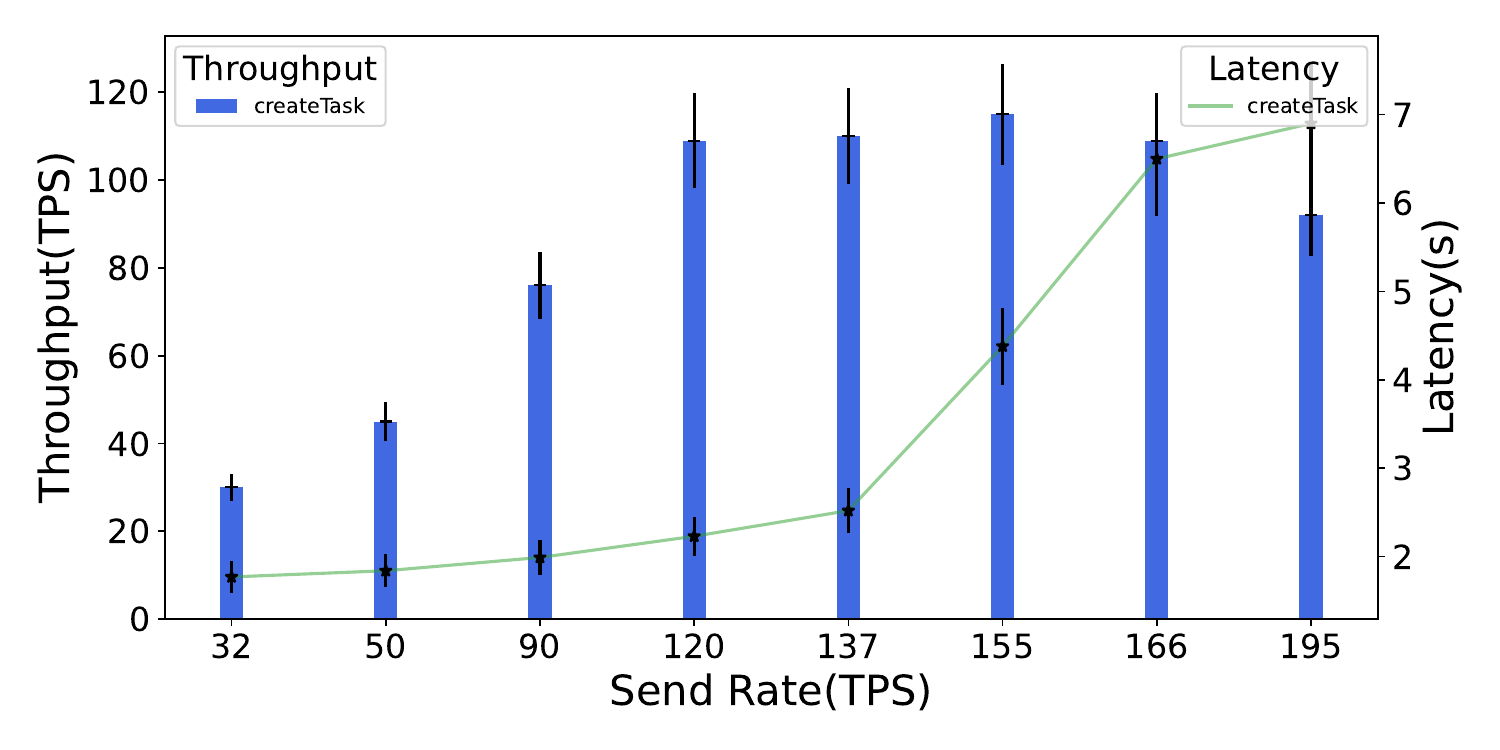}
        \label{fig3a}
    }
    \subfloat[Task Subscription ]{
        \hspace{-0.2in}
        \includegraphics[scale=0.23]{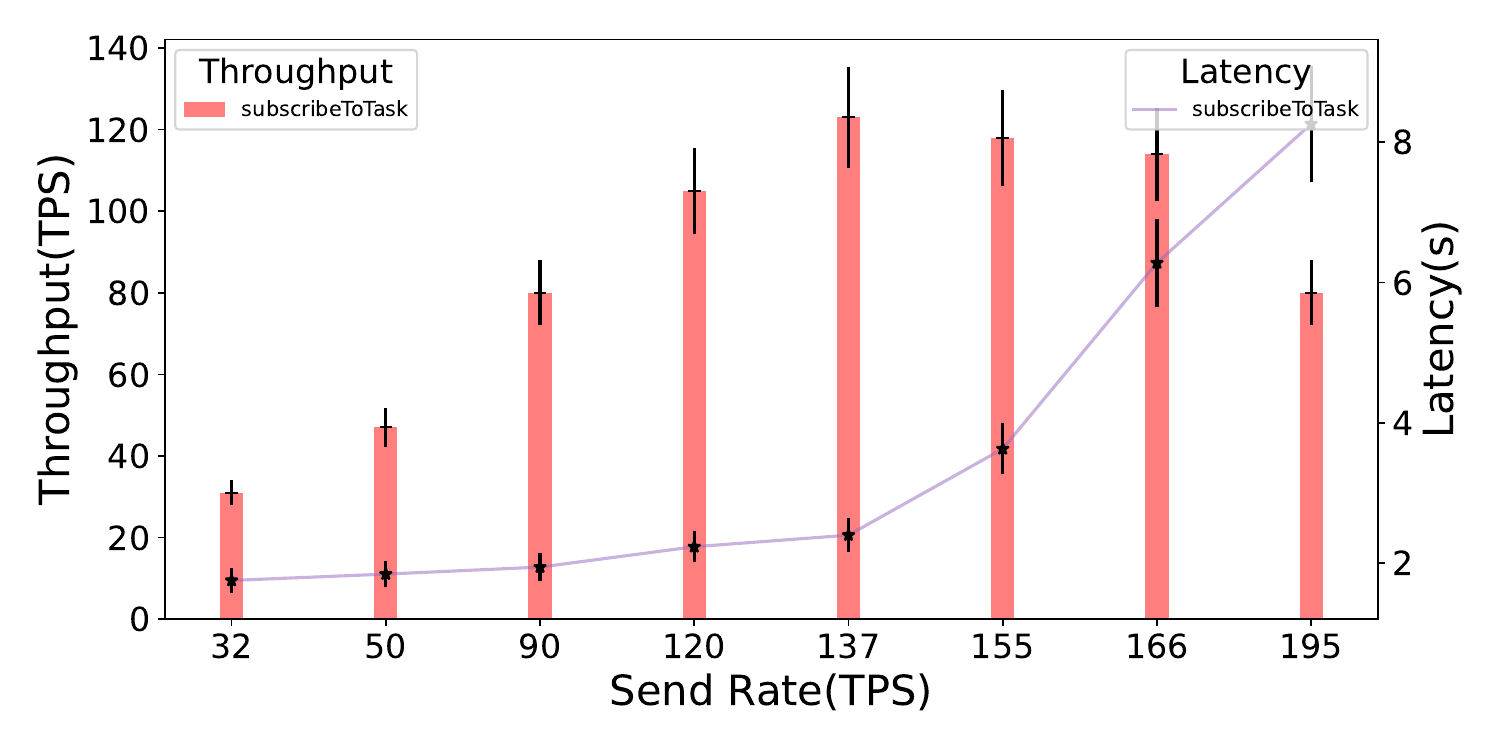}
        \label{fig3b}
    }
    \subfloat[Model Submission ]{
    \hspace{-0.2in}
        \includegraphics[scale=0.23]{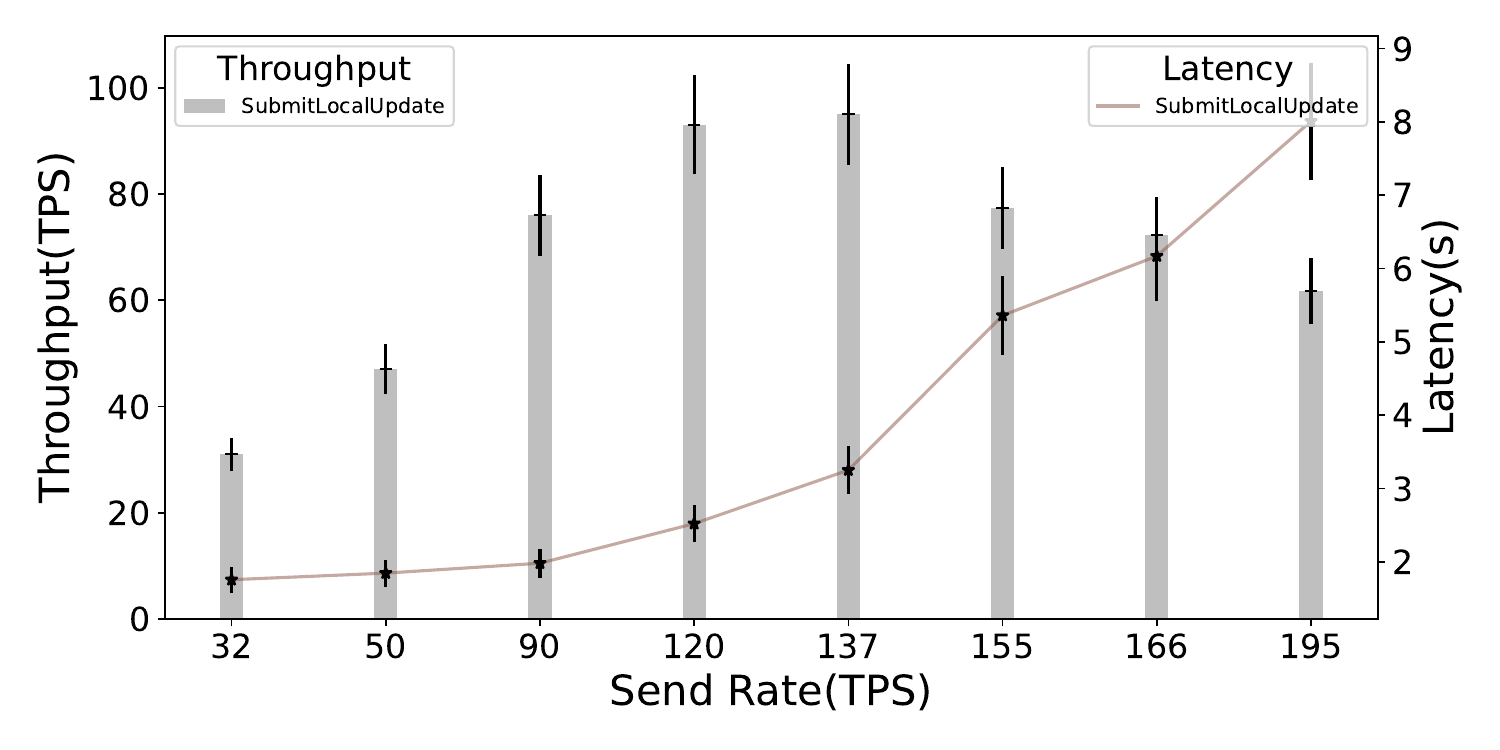}
        \label{fig3c}
    }
    \caption{Latency and Throughput of VerifBFL}
    \label{fig3}
\end{figure*}

Fig.\ref{fig3} presents a summary of the results obtained by invoking the system's functions with different send rates;

The performance evaluation presented in Fig. \ref{fig3a} illustrates the changes in throughput and latency as the \textit{createTask} function is invoked with varying transaction sending rates. The sending rate, representing the number of transactions emitted per second (TPS), corresponds to the frequency of function calls per second at the user level. As the transaction sending rate increases, throughput similarly rises, peaking at approximately 120 TPS at a sending rate of 155 TPS. During this period, latency remains low, indicating that the system is operating efficiently without network congestion. However, once the system reaches its performance threshold, the latency escalates sharply, highlighting a saturation point.

The \textit{subscribeToTask} function follows a comparable trend (Fig. \ref{fig3b}), achieving a maximum throughput exceeding 120 TPS with sending rates around 137 TPS. However, the \textit{submitLocalUpdate} function demonstrates a slightly lower peak throughput (Fig. \ref{fig3c}). This reduction is attributed to the computational overhead introduced by offloading proof verification to the decentralized oracle network (DON). Each invocation of this function generates two transactions, hence the observed performance difference. Nevertheless, the cost of integrating proofs into our system is quite transparent. The approach we propose effectively balances security and efficiency, with SNARK proofs contributing to the verifiability and integrity of the process without imposing prohibitive computational costs.

\section{Conclusion}
\label{sec:conclusion}
\quad In this paper, we proposed VerifBFL, a novel verifiable blockchain-based federated learning framework for crowdsourcing. VerifBFL integrates blockchain technology and cryptographic protocols to build a trustless, privacy-preserving, and verifiable system for federated learning. The verifiability of both the local training and aggregation processes is ensured through the employment of zk-SNARKs and incrementally verifiable computation (IVC). The proofs of training and aggregation are verified on the blockchain, thereby ensuring the integrity and auditability of each participant's contributions. Furthermore, to safeguard training data from inference attacks, we employ differential privacy. Finally, to assess the efficacy of the proposed protocols, we constructed a proof of concept utilizing emerging tools. The results show that generating proofs for local training and aggregation in VerifBFL takes around 81s and 2s respectively, while verifying them on the chain takes less than 0.6s.

\quad Despite the framework's success in maintaining system verifiability and accountability, the performance results underscore areas for improvement, particularly in scalability. While our approach effectively balances security and efficiency, future work should focus on optimizing performance under high-load conditions to further enhance the system’s scalability.



\bibliographystyle{IEEEtran}
\bibliography{ref}

\end{document}